\documentclass[conference]{IEEEtran}
\IEEEoverridecommandlockouts
\usepackage{cite}
\ifCLASSINFOpdf
  \usepackage[pdftex]{graphicx}
\else
\fi
\usepackage[tight,footnotesize]{subfigure}
\usepackage{url}

\begin{document}
\title{Measurement and Evaluation of ENUM Server Performance}

\author{\IEEEauthorblockN{Charles Shen and Henning Schulzrinne}
\IEEEauthorblockA{Department of Computer Science, Columbia University\\
1214 Amsterdam Avenue, New York, New York 10027\\
Email: \{charles, hgs\}@cs.columbia.edu}

}

\maketitle

\begin{abstract}
ENUM is a DNS-based protocol standard for mapping E.164 telephone numbers to Internet Uniform Resource Identifiers (URIs). It places unique requirements on the existing DNS infrastructure, such as data scalability, query throughput, response time, and database update rates. This paper measures and evaluates the performance of existing name server implementation as ENUM servers. We compared PowerDNS (PDNS), BIND and Navitas. Results show that BIND is not suitable for ENUM due to its poor scaling property. Both PDNS and Navitas can serve ENUM. However, Navitas turns out to be highly optimized and clearly outperforms PDNS in all aspects we have tested. We also instrumented the PDNS server to identify its performance bottleneck and investigated ways to improve it.
\end{abstract}
\IEEEpeerreviewmaketitle

\section{Introduction} \label{sec:intro}

ENUM~\cite{rfc3761} is a protocol standard developed by the Internet Engineering Task Force (IETF) to translate global Public Switched Telephone Network (PSTN) phone numbers, also known as E.164 numbers, into Internet URIs. It serves as a bridge in the ongoing convergence of traditional telecommunications services and Internet services. ENUM is built upon the existing DNS infrastructure. However, the special data characteristics and service quality expectation of ENUM impose unique requirements on existing DNS software operating as an ENUM server. These requirements include accommodation of a huge number of records, providing high query throughput for both existing records and non-existing records in the database within a tight response time budget, and allowing background bulk database update without severely affecting the frontend query performance. There has been no comprehensive study on whether and how all these requirements can be satisfied by name servers that people commonly use today. Our work addresses this issue.

In this paper we report on the development of an ENUM benchmarking tool called enumperf based on the widely used DNS performance measurement software queryperf ~\cite{queryperf}. Using this tool, we derived and analyzed a series of test results for ENUM servers with three different name server implementations, BIND~\cite{bind}, PDNS~\cite{pdns} and Navitas~\cite{navitas}. BIND is the traditional open source DNS server that stores domain name information in text format zone files. It still dominates today's DNS deployment. PDNS, also open source, supports not only zone files, but also most of the common database backends such as MySQL~\cite{mysql}. It has become one of the most popular BIND substitutes. Navitas is a commercial name server produced by Nominum Inc. dedicated for ENUM application. It supports Nominum's own database drivers and Berkeley DB. Among the three servers, PDNS is more modern compared with BIND and unlike Navitas is open source. Therefore we focused on PDNS for a more thorough analysis which involves instrumenting detailed server profiling to understand the contribution of various system components to its overall performance. BIND and Navitas are generally treated as black boxes in our study. We contrasted the performance of these three servers in aspects including query throughput, response time, database scaling, server processor capability and server memory usage. Due to space constraints, comparison on server processor capability and server memory usage is not included in this paper. Interested readers are referred to our technical report~\cite{shen06enumTR} for those results as well as greater details on items contained in this paper.

Our tests show that, on our platform, the ENUM query throughput is of much larger concern than query response time. BIND is especially poor in throughput for querying existing records, which disqualifies it as a serious ENUM server. PDNS can operate reasonably well as an ENUM server at the scale of 5M database records. However it suffers a critical bottleneck on serving non-existing records because of excessive number of database lookups. This makes it less appealing in current ENUM deployment where queries for non-existing records are expected to be fairly common. We also illustrate how this problem can be alleviated by appropriate PDNS caching. Finally, Navitas turns out to perform significantly better than the other two servers in all aspects we have tested. This shows that an optimized implementation can have remarkable impact on specific server performance.

The rest of the paper is organized as follows: Section \ref{sec:expsetup} describes our measurement setup and experimental methodology. Section \ref{sec:pdnsperf} presents detailed PDNS performance evaluation. Section \ref{sec:perfcomp} gives a summarized comparison by contrasting the PDNS performance with our results obtained from BIND and Navitas. Section \ref{sec:related} discusses related work. Concluding remarks are offered in Section \ref{sec:conclusion}.

\section{Experimental Setup and Methodology} \label{sec:expsetup}

\subsection{Hardware Profile}

Our test platform consists of three Linux machines all running Red Hat Enterprise 3.4.4 with kernel version 2.6.9. Two of them have four 3\,GHz Intel Xeon CPUs, 8\,GB memory, and 150\,GB hard drive each. The third one has two 3\,GHz Intel Xeon CPUs, 2\,GB memory, and 80\,GB hard drive. The three machines are connected via 100M Ethernet connections. We checked to make sure that the network speed is not a bottleneck during the benchmarking.

\subsection{ENUM Server}

We evaluated the performance of BIND, PDNS and Navitas as ENUM servers. Our choices are not meant to deprecate any other existing name server implementations. Instead we see these three servers as representing a mix of typical factors such as large user population, traditional and modern flavor, open source and commercial product, multifunctional and dedicated authoritative DNS architecture.

We used PDNS release 2.9.19 with MySQL version 4.1.12 as its database backend. In the rest of this document, the term PDNS is used for both the combined PDNS-MySQL server, and its PDNS component only, depending on the context. We chose InnoDB as the MySQL storage engine. The maximum memory allocated to MySQL is half of the server physical memory or 4\,GB. The number of concurrent InnoDB threads is set to five, which is the sum of the number of processors and hard disk drives in the server system as recommended by the MySQL manual. We kept most of the default configuration options for PDNS unchanged but disabled the recursive option and the wildcard records option. The number of PDNS handling threads is set to be equal to the number of MySQL InnoDB threads, unless specified otherwise. Our tests by default put PDNS and MySQL server in the same machine although we put the two in different machines when testing performance limitation caused by processor capability. The two server placement settings are referred to as collocated and non-collocated PDNS server architecture, respectively.

The BIND version we tested is 9.2.4. We disabled the BIND recursion function. The Navitas version we tested is 2.6.0.0. Navitas is designed for ENUM so we used its default settings including its {\it enum\_mnpz} database driver and a 100\,MB cache size.

We do a complete database table scan before our tests involving server caching to minimize the effect of initial cache heat-up.

\subsection{ENUM Clients and Enumperf Test-suite}

We extend the Nominum modified queryperf version 2.1 and use it as the ENUM client. Our extension enables queryperf to generate random queries on the fly within a predefined domain name range. We also reduce the number of clients each queryperf process can simulate from its default value of 20 to 1 because the default setting can too easily saturate a server. In our tests we step up the load by increasing the number of queryperf processes, which is also the number of ENUM clients. Note that queryperf is a closed-loop load generator. This has two important implications. First, the actual query load generated with equal number of clients is usually not the same under different server configurations. Second, as the server approaches its saturation point, continuing to increase number of clients can result in the subsequent total load generated fluctuating around the server maximum load capacity because the server working condition may not be completely stable at that point.

\begin{figure}[htp]
\centerline{
\includegraphics[width=6cm]{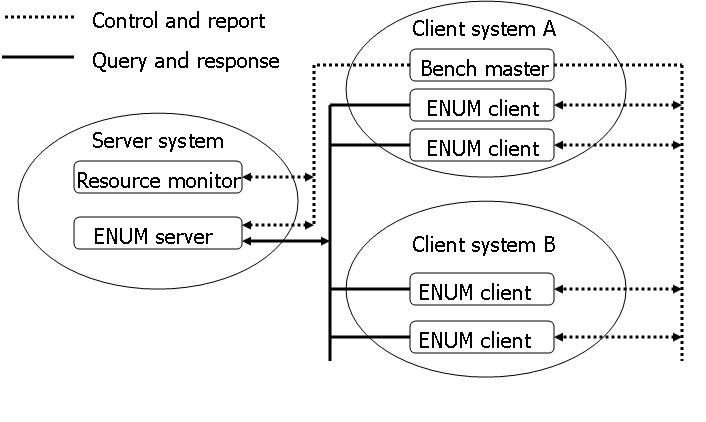}
\vspace{-0.25in}
}
\caption{Enumperf testbed architecture}\label{fig:enumperf.png}
\vspace{-0.2in}
\end{figure}

Based on our extended queryperf, we developed the enumperf test-suite. It consists of a set of perl and shell scripts running on the Linux platform and uses the Unix program {\it rexec} for executing processes on remote machines, modelled on the Session Initiation Protocol (SIP) server benchmarking tool SIPStone~\cite{sipstone}. On our three-machine testbed, we deploy enumperf as shown in Fig.~\ref{fig:enumperf.png}. The ENUM server is running on one of the 4-CPU machines denoted as server system. The other two machines run as client systems. One of the client systems runs the Bench Master script. Execution of the Bench Master script causes its host system to connect to appropriate client and server systems to start resource monitoring and create the desired number of ENUM clients at pre-defined intervals during each test step. When all test steps are finished, the Bench Master collects all results data from client and server systems and processes them to produce statistics. In the cases where PDNS and its database backend need to be on separate machines, we use both of the 4-CPU machines as server systems.

\vspace{-0.1in}

\subsection{ENUM Record Sets and Test Query Space}

An ENUM record maps a Fully Qualified Domain Name (FQDN) derived from a telephone number to a corresponding URI. The phone number to FQDN conversion is done by reversing the digits of the phone number, inserting dots between each, and appending the appropriate domain name at the end. An example record is as follows:

{\small \verb#$ORIGIN 4.3.2.1.9.3.9.2.1.2.1.e164.arpa. #

\verb#    IN NAPTR 0 0 "u" "E2U+sip" \ #

\verb#    "!^.*$!sip:info@enum.example.com!" #}

The FQDN {\small \verb#4.3.2.1.9.3.9.2.1.2.1.e164.arpa.#} corresponds to a typical US phone number {\small \verb#+1-212-939-1234#}. {\small \verb#IN#} stands for {\it Internet}, {\small \verb#NAPTR#} indicates that this is of DNS {\it NAPTR} record type. The two {\small\verb+0+}s represent order and preference in using the record. The {\small\verb+u+} field says the record is terminal and the output is a URI. {\small\verb#E2U+sip#} is the service type field. {\small\verb#!^.*$!sip:info@enum.example.com!#} is a regular expression that produces {\small\verb#info@enum.example.com#} as the resulting URI for the FQDN.

We generated three record sets at the size of 500k, 5M and 20M, respectively. The majority of our tests used the 5M-record set. The other two sets are used for scaling tests.

We measured query performance for both existing records and non-existing records. Tests for existing records use all records in the server as the query space. Tests for non-existing records are divided into two cases. First, the server is authoritative for a subdomain of those non-existing records being queried. This is referred to as query for non-existing authoritative records. Second, the server is not authoritative for any of the subdomains of those non-existing records being queried. This is referred to as query for non-existing non-authoritative records. The two types of non-existing records will only be distinguished in our initial series of tests. Then we will focus only on non-existing non-authoritative records which is the type that is expected to be much more common.

\section{PDNS Performance Evaluation} \label{sec:pdnsperf}

\subsection{Overall Query Performance} \label{sec:overallperf}

\begin{figure}[htp]
\begin{center}
\vspace{-0.25in}
\subfigure[Throughput]{\label{fig:i1-thruput.png}
\includegraphics[width=4.25cm]{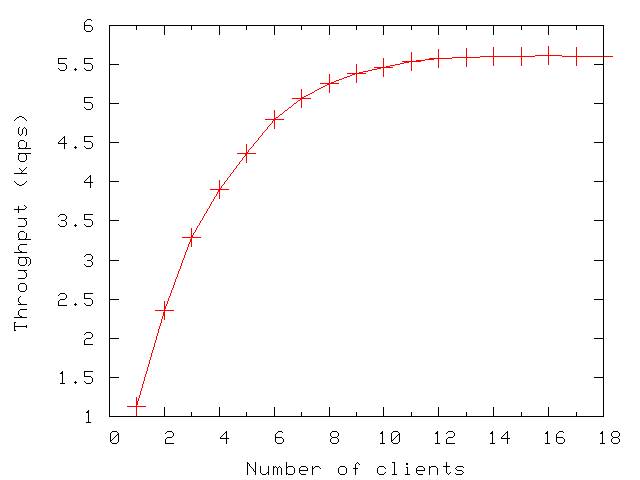}}
\vspace{-0.15in}
\subfigure[Response time]{\label{fig:i1-delay.png}
\includegraphics[width=4.25cm]{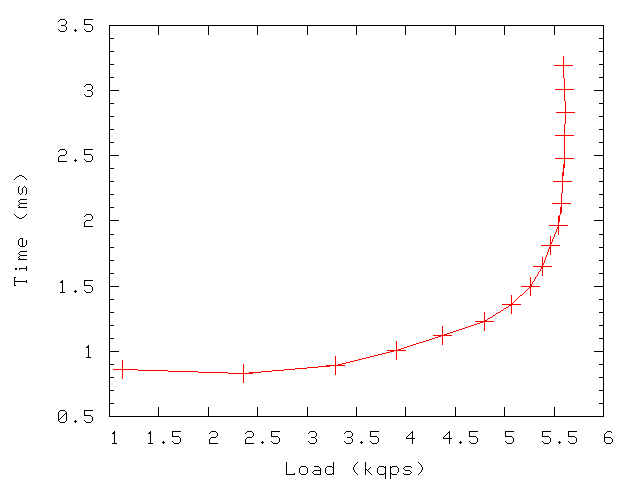}}
\end{center}
\caption{PDNS performance for existing records with 5M-record set}\label{fig:i1}
\vspace{-0.3in}
\end{figure}

\begin{figure}[htp]
\begin{center}
\subfigure[Throughput]{\label{fig:i1ne-nea-thruput.png}
\includegraphics[width=4.25cm]{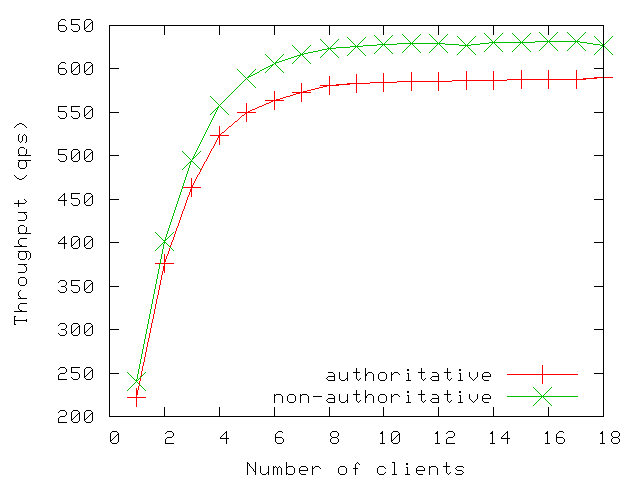}}
\vspace{-0.15in}
\subfigure[Response time]{\label{fig:i1ne-nea-delay.png}
\includegraphics[width=4.25cm]{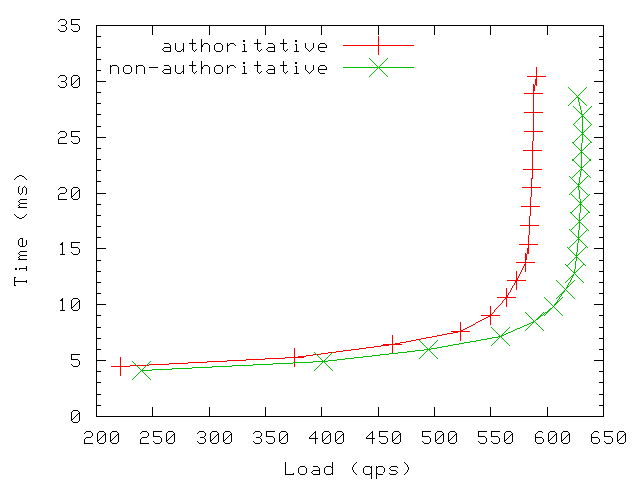}}
\end{center}
\caption{PDNS performance for non-existing records with 5M-record set}\label{fig:i1ne}
\vspace{-0.1in}
\end{figure}

The overall PDNS performance is obtained on the collocated PDNS architecture with the 5M-record set loaded in the server. Fig.~\ref{fig:i1} illustrates the results for querying existing records. We see that PDNS throughput grows almost linearly from $1,\,120~$queries per second (qps) to $3,300\,~$qps as the number of clients increases from one to three. Within this throughput range the response time is below $0.8\,~$ms. When the number of clients continue to increase, the throughput growth rate slows down and the response time starts to rise. Finally the throughput saturates at around $5,\,500\,~$qps with response time rising dramatically after the number of clients reaches $12$.

PDNS performance for querying non-existing records, as shown in Fig.~\ref{fig:i1ne}, is significantly worse than that for existing records. The peak throughput for non-existing records is only one ninth of that for existing records, and the response time is $7.5~$times longer. In addition, querying non-existing non-authoritative records achieves about $80~$qps higher saturation throughput and slightly lower response time compared to querying non-existing authoritative records.

\subsection{Analysis of Response Time Components}

We examined PDNS source code for its internal architecture and the query processing rules. As a result, the PDNS server processing time for a specific query is found to consist of the following main components:
\begin{enumerate}
\item server queueing time. This is the period of time an incoming query spends  in the queue waiting for a PDNS query handling thread to be available.
\item ``make canonic'' database lookup time. This lookup checks the canonical name record of a queried domain name in case the submitted domain name is an alias.

\item ``actual type'' database lookup time. This lookup checks the actual type of record ({\it NAPTR} in case of ENUM) for the queried domain.

\item Start of Authority (SOA) related database lookup time. An SOA record indicates that the server is authoritative for a specified domain. In this step the server checks SOA records of all subdomains of the queried domain. If such records are found, the server may invoke more lookups for the name server information of those specific subdomains.
\item other server processing time. This component contains all server processing other than those listed above.
\end{enumerate}

Specifically, querying existing records involves three of the above components, namely server queueing, ``make canonic'' database lookup, and other server processing time; while querying non-existing records involves all these five components.

\begin{figure}[htp]
\centerline{
\includegraphics[width=4.25cm]{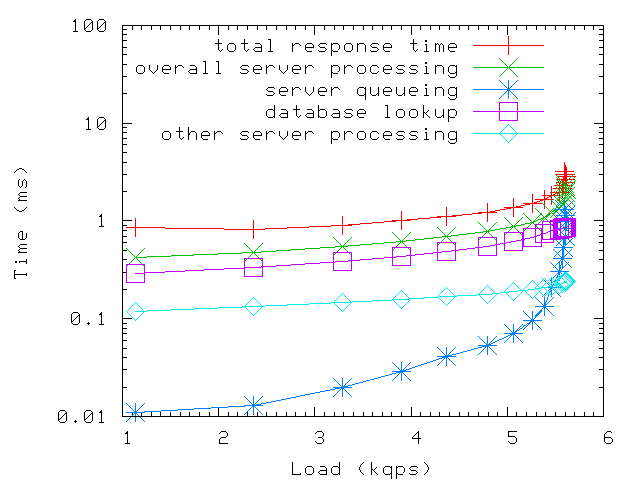}
\vspace{-0.15in}
}
\caption{PDNS response time components for querying existing records with 5M-record set}\label{fig:i1-compdelay.png}
\vspace{-0.1in}
\end{figure}

We instrumented PDNS server code and quantified the various time components. The response time components for querying existing records are shown in Fig.~\ref{fig:i1-compdelay.png}. The total response time in the figure corresponds to the same value in Fig.~\ref{fig:i1-delay.png}, which is the client side measured response time. The small difference between total response time and overall server processing time is mainly the latency when the query packets stay in the server system receiving buffer before being delivered to the ENUM server process. The overall server processing time consists of the three components we mentioned above. We see that before the server is saturated, the database lookup time is the dominant factor and accounts for over $60\%$ of the total server processing time. After reaching the saturation point, this value is capped at around $0.85~$ms because the load PDNS can deliver to the database backend has reached the maximum. The queuing delay component is initially small. But it grows rapidly as the server approaches the saturation point, after which it replaces the database lookup time as the main factor in the overall server processing time. The ``other server processing time'' latency component remains relatively constant and small all the time, with an overall increasing trend.

Response time components figures were also obtained for querying non-existing records. It is found that excessive number of SOA related database lookups caused the severe PDNS performance degradation for non-existing records compared to that of existing records. The number of these lookups could be more for non-existing authoritative records than for non-existing non-authoritative records, making the performance of the former worse than that of the latter. Nevertheless, the performance differences between the two types of non-existing records is not remarkable, so we will only look at the more common non-authoritative records when we evaluate performance for non-existing records in the rest of the paper.

\subsection{Performance Limited by Database Scaling}

We consider the scaling of database record set in this section. Our main record set contains 5M records, which are all loaded in MySQL memory. We tested two additional record sets, one is a scaling down of 500k records, also fully loaded in MySQL memory; the other is a scaling up of 20M records. In the 20M-record set case all records cannot fit into the 4G memory we allocated to MySQL.

\begin{figure}[htp]
\begin{center}
\vspace{-0.1in}
\subfigure[Throughput]{\label{fig:i1-l1-u1-thruput.png}
\includegraphics[width=4.25cm]{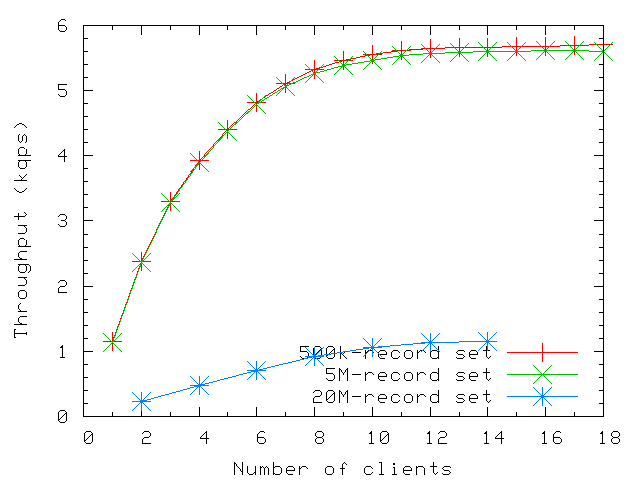}}
\subfigure[Response time]{\label{fig:i1-l1-u1-delay.png}
\includegraphics[width=4.25cm]{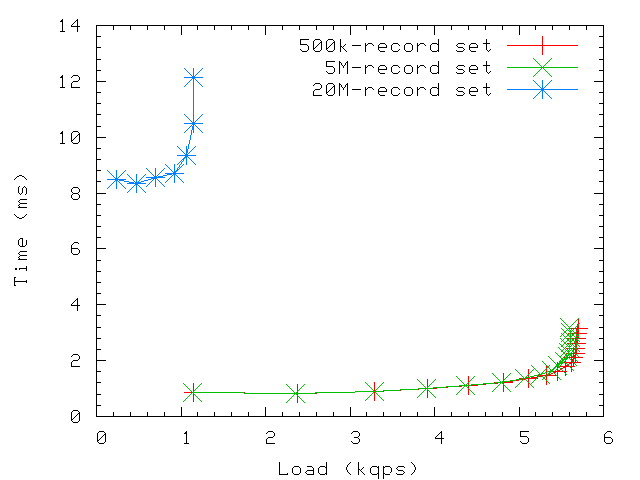}}
\end{center}
\vspace{-0.15in}
\caption{Comparison of PDNS performance for existing records with different record set size}\label{fig:liu1}
\end{figure}

\begin{figure}[htp]
\begin{center}
\subfigure[Throughput]{\label{fig:i1ne-l1ne-thruput.png}
\includegraphics[width=4.25cm]{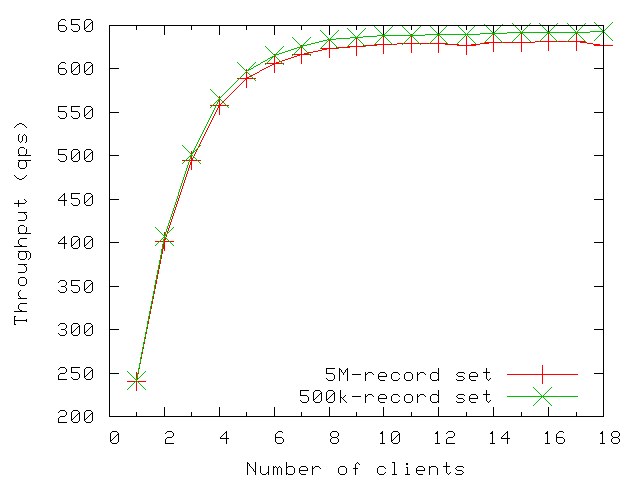}}
\subfigure[Response time]{\label{fig:i1ne-l1ne-delay.png}
\includegraphics[width=4.25cm]{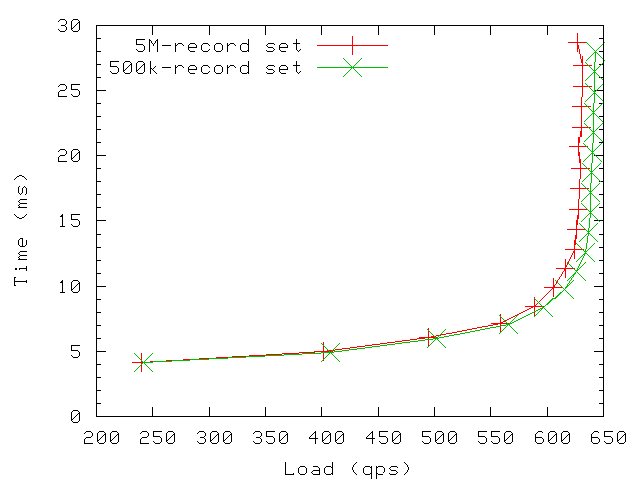}}
\end{center}
\vspace{-0.15in}
\caption{Comparison of PDNS performance for non-existing records with different record set size}\label{fig:li1ne}
\vspace{-0.05in}
\end{figure}

Figure \ref{fig:liu1} shows the PDNS throughput and response time for existing records with varying database sizes. When the records are in memory, as in the cases of 500k and 5M record set, the response time is almost constant and the maximum throughput differs only by $2\%$. The comparison of performance for querying non-existing records with 500k and 5M record sets is similar, as shown in Figure \ref{fig:li1ne}.

Figure \ref{fig:liu1} also shows, however, the performance degrades significantly when records are out of memory. The maximum throughput for the 20M-record set is reduced by over $80\%$ compared with the other two, and response time is increased by over seven times.

\vspace{-0.05in}
\subsection{Improving Performance by PDNS Caching}

The tests on PDNS in the previous sections assume the basic operation mode without any PDNS caching enabled. In this section, we look at the PDNS performance in caching mode. PDNS provides two levels of caching, packet caching and query caching. The former caches the entire query and response packets. The latter caches returned database query results. Both query cache and packet cache can be applied to positive and negative queries. A specific type of cache is enabled by setting the corresponding Time To Live (TTL) value to be positive. In our tests we set the TTL value to be longer than the experiment duration to obtain the performance under a full cache scenario.

\begin{figure}[htp]
\begin{center}
\vspace{-0.1in}
\subfigure[Throughput]{\label{fig:i1-i2-ncl-i4-ncl-thruput.png}
\includegraphics[width=4.25cm]{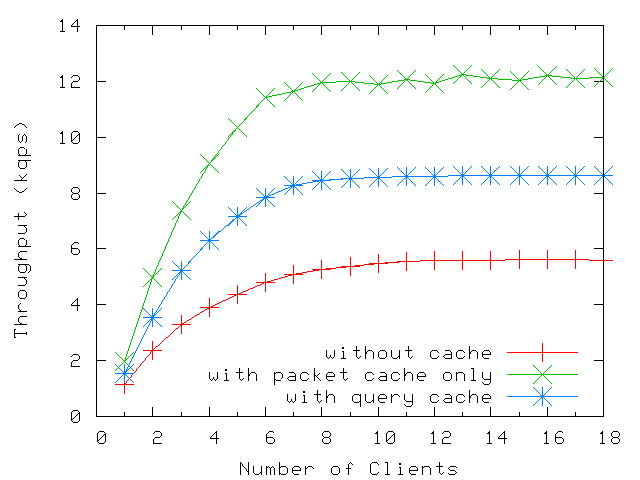}}
\subfigure[Response time]{\label{fig:i1-i2-ncl-i4-ncl-delay.png}
\includegraphics[width=4.25cm]{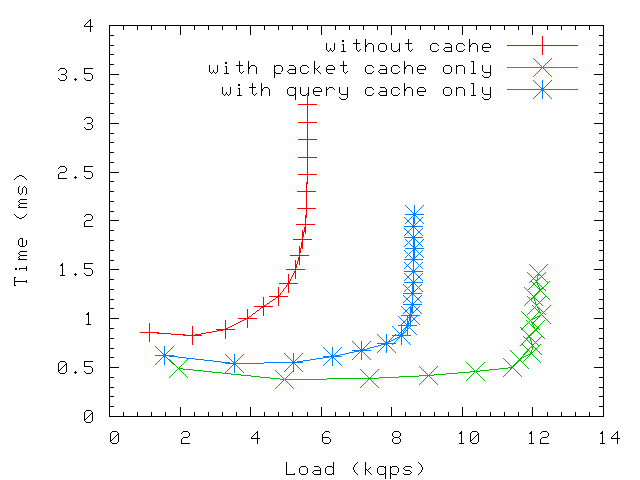}}
\end{center}
\vspace{-0.15in}
\caption{Comparison of PDNS performance with and without caching for querying existing records with 5M-record set, after disabling explicit cache maintenance}\label{fig:i124ncl.png}
\vspace{-0.15in}
\end{figure}

During our caching tests and server profiling we discovered an implementation flaw with PDNS cache maintenance. Specifically, the PDNS server does an explicit cache check to clean up expired cache entries after every 1,000 queries are processed. This is an $O(n)$ operation. When the cache size is large, the cost of this task eliminates the benefit of caching and even makes the performance worse than that without caching. It is not difficult to fix this issue. Therefore the caching performance results we presented in Fig.~\ref{fig:i124ncl.png} are obtained after disabling this explicit maintenance routine. We see that the throughput increases by $100\%$ more with packet cache enabled, or $50\%$ more with query cache enabled. The packet cache is more efficient than query cache for existing records. But the query cache can be especially useful to alleviate the performance bottleneck for non-existing records because it can cache results for common subdomain SOA related lookups for non-existing records. We repeated the test for non-existing records in the 5M-record set with negative query cache enabled and found the resulting maximum throughput to be $7,000~$qps, which is an improvement of an order of magnitude. Note that cached SOA lookup is not the only reason for this improvement. There could also be cached direct query lookups. The characteristics of negative query space has an impact as well.

\subsection{Performance under Database Update Load} \label{sec:pdnsupdate}

In our PDNS server, the update load is applied to the MySQL database. We considered two types of background loads, ADD and UPDATE. ADD load uses the MySQL {\it insert} command to add new records; UPDATE load uses the MySQL {\it update} command to modify existing records in the database. The results comparing the performance with and without background load show that the maximum query throughput is reduced by roughly $20\%$, and the response time is increased by up to a few hundred microseconds with the background load.

In another test, we kept the number of clients at three, which generates a query load within the server's normal operation range. At the same time, we initiated updates of $10,000$ different existing records $20~$ times at every $30$-second interval. At the end of the test we found that PDNS achieves a $3,000~$qps throughput, and updating $10,000$ records takes on average $2.7~$s. This corresponds to an update rate of $3,700$ records per second.

On the whole, PDNS query performance remains at a reasonable level while background database update process achieving update rate of several thousands records per second.

\section{Performance Comparison of PDNS, BIND and Navitas} \label{sec:perfcomp}

This section offers a summarized comparison by contrasting the PDNS performance presented above with our test results obtained with BIND and Navitas.

\subsection{Performance for Querying Existing Records}

\begin{figure}[htp]
\begin{center}
\vspace{-0.15in}
\subfigure[Throughput]{\label{fig:i2ncl-w1-mnpz-thruput.png}
\includegraphics[width=4.25cm]{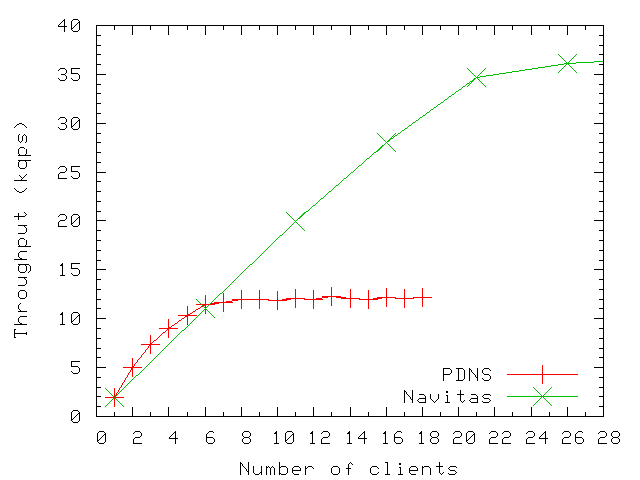}}
\subfigure[Response time]{\label{fig:i2ncl-w1-mnpz-delay.png}
\includegraphics[width=4.25cm]{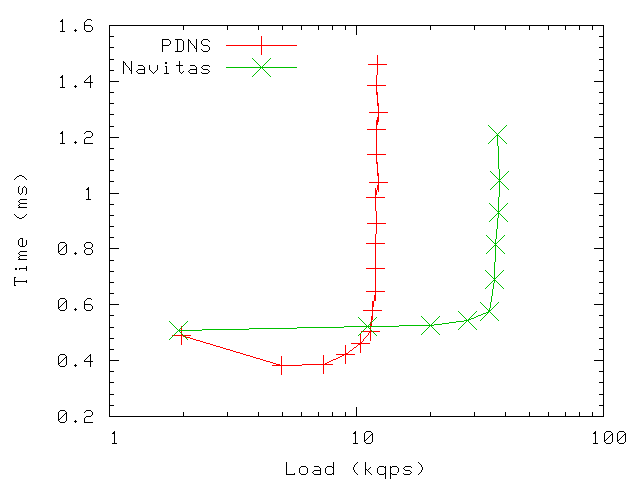}}
\end{center}
\vspace{-0.1in}
\caption{Performance comparison of PDNS and Navitas for existing records with 5M-record set}\label{fig:iw}
\vspace{-0.15in}
\end{figure}

Comparison of PDNS and Navitas performance for querying existing records using the 5M-record set is shown in Fig.~\ref{fig:iw}. The value for PDNS represents its overall best performance when full packet cache is enabled and explicit cache cleanup disabled. Navitas could be using a different caching mechanism. Its default setting gives an 85\% to 90\% cache hit rate at the maximum throughput for our 5M-record test. We see from Fig.~\ref{fig:i2ncl-w1-mnpz-thruput.png} that the maximum throughput of Navitas, about $39,000~$qps, is more than three times that of PDNS. The BIND performance is not included in this figure because it is only $1/50$ of the Navitas throughput.

As far as response time is concerned, Fig.~\ref{fig:i2ncl-w1-mnpz-delay.png} shows that Navitas and PDNS offer a similar, relatively constant value below $0.5~$ms within their respective load capacity. To understand whether this is sufficient to meet the service quality expectation, we can compare it with its allowed delay budget. Since ENUM is primarily used for destination address lookup as part of the Voice over IP (VoIP) call signaling process, assuming people expect at least as good quality in VoIP as in PSTN, one can define the ENUM lookup response time budget following similar operations in making a PSTN call. This could be, for example, the routing function in a PSTN switch when processing ISDN User Part (ISUP) messages. The mean value of total switch response time budget for such a message is about $200~\,$ms~\cite{gr1364}. Our response time obtained from both PDNS and Navitas is clearly well-below this range. In fact, these delay values of a few milliseconds are unlikely to be a concern if we further consider the several-second-long normal PSTN post dial delay.

\subsection{Performance for Querying Non-existing Records}

\begin{figure}[htp]
\begin{center}
\vspace{-0.1in}
\subfigure[Throughput]{\label{fig:i1neNncl-o1ne-w1ne-mnpz-thruput.png}
\includegraphics[width=4.25cm]{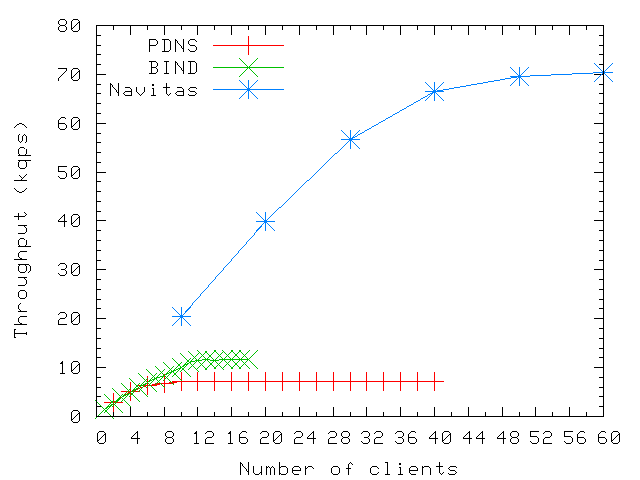}}
\subfigure[Response time]{\label{fig:i1neNncl-o1ne-w1ne-mnpz-delay.png}
\includegraphics[width=4.25cm]{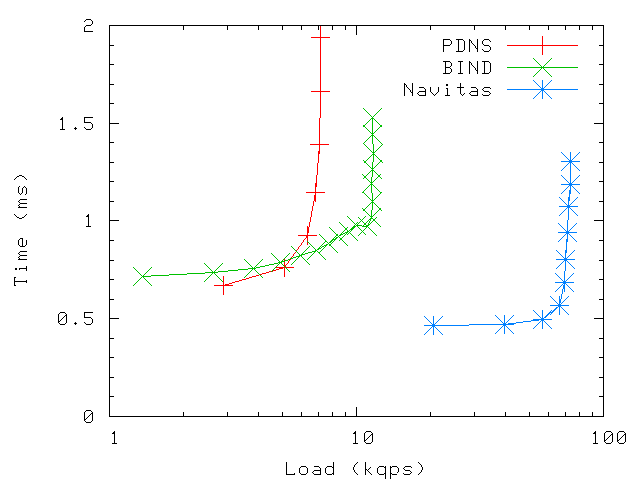}}
\end{center}
\vspace{-0.1in}
\caption{Performance comparison of PDNS, BIND and Navitas for non-existing records with 5M-record set}\label{fig:iowne}
\vspace{-0.1in}
\end{figure}

Comparison of PDNS, BIND and Navitas performance for querying non-existing records using the 5M-record set is shown in Fig.~\ref{fig:iowne}. The value for PDNS represents its best performance when negative query cache is enabled and explicit cache cleanup disabled. Even in this case, Fig.~\ref{fig:i1neNncl-o1ne-w1ne-mnpz-thruput.png} illustrates that the maximum Navitas throughput of $72,000$~qps is approximately 10 times that of PDNS and 6 times that of BIND. Note that for non-existing records, BIND outperforms PDNS in throughput. In terms of response time, Fig.~\ref{fig:i1neNncl-o1ne-w1ne-mnpz-delay.png} shows that Navitas has the shortest value, which is below $0.5~\,$ms within load capacity. The response time of PDNS and BIND are similar, both at a few hundred microseconds more than that of Navitas. All total response time under normal server operation is also well within the allowed budget.

\subsection{Performance Limited by Database Scaling}

\begin{figure}[htp]
\begin{center}
\vspace{-0.2in}
\subfigure[Throughput]{\label{fig:n-o-thruput.png}
\includegraphics[width=4.25cm]{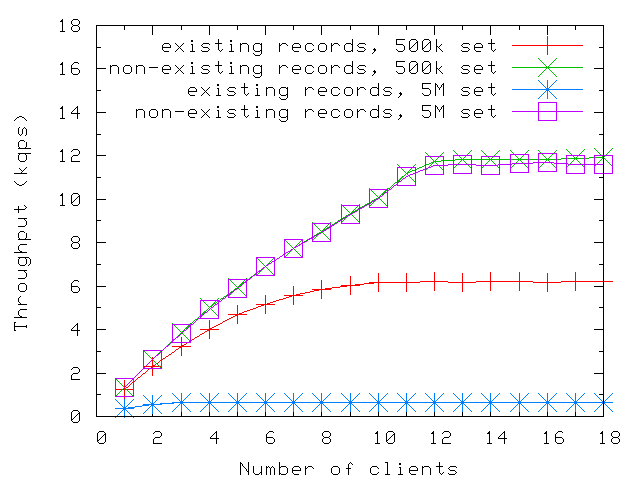}}
\subfigure[Response time]{\label{fig:n-o-delay.png}
\includegraphics[width=4.25cm]{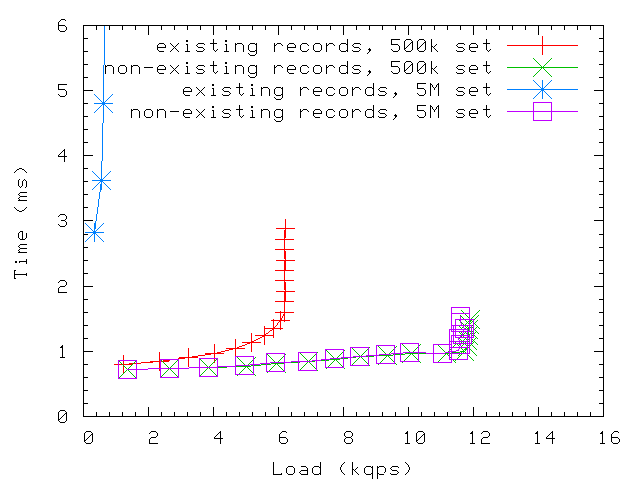}}
\end{center}
\vspace{-0.15in}
\caption{Comparison of BIND performance scaling}\label{fig:no}
\vspace{-0.1in}
\end{figure}

\begin{figure}[htp]
\begin{center}
\subfigure[Throughput]{\label{fig:y-w-z-mnpz-thruput.png}
\includegraphics[width=4.25cm]{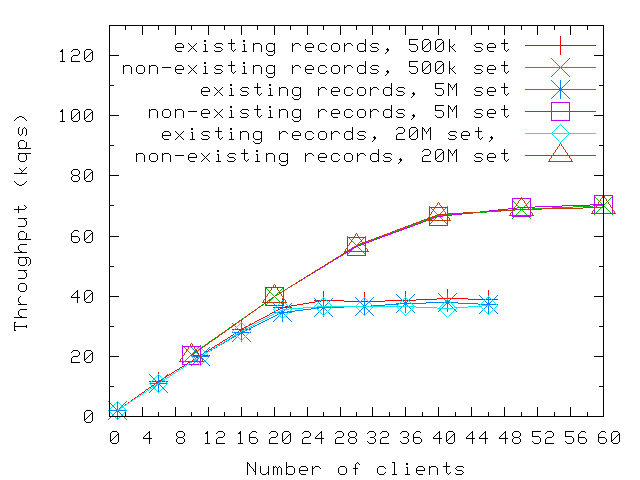}}
\subfigure[Response time]{\label{fig:y-w-z-mnpz-delay.png}
\includegraphics[width=4.25cm]{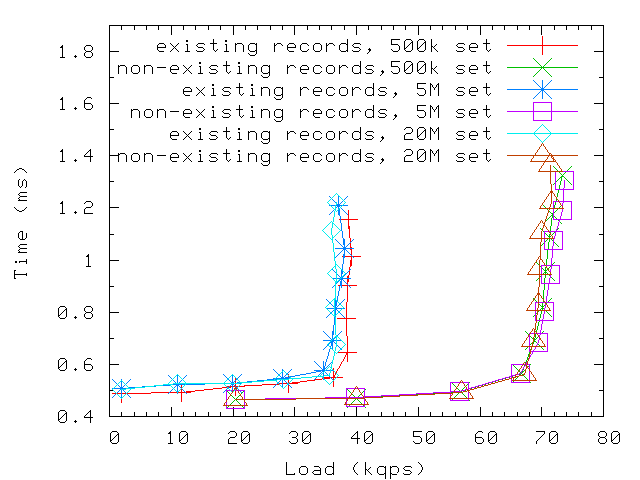}}
\end{center}
\vspace{-0.15in}
\caption{Comparison of Navitas performance scaling}\label{fig:ywz}
\vspace{-0.25in}
\end{figure}

The database scaling property of PDNS is shown earlier in Fig.~\ref{fig:liu1} and Fig.~\ref{fig:li1ne}. We present BIND scaling property in Fig.~\ref{fig:no} and Navitas scaling property in Fig.~\ref{fig:ywz}. In general, when records fit into memory, Navitas offers the best scaling property among the three servers. PDNS scales well mainly for existing records. BIND scales poorly for existing records but displays good scaling property for non-existing records. When all records do not fit into memory, BIND will fail to work, while PDNS performance degrades dramatically. Navitas should still work according to its manual, although we have not tested the performance difference.

\vspace{-0.05in}
\subsection{Performance under Database Update Load}

While the maximum PDNS throughput appears to drop by about 20\% at the presence of background update load as mentioned in Section~\ref{sec:pdnsupdate}, similar tests show that the maximum throughput of Navitas is largely unaffected by background update load. On the other hand, the variation of response time under background update load is within sub-millisecond level for both PDNS and Navitas servers. BIND is not tested because it requires a zone reloading for updates and cannot answer queries during the reloading.

\begin{table}[th]
\vspace{-0.1in}
\begin{center}
\begin{tabular}{|l|l|l|l|l|}
\hline
Parameter & Navitas & PDNS \\
\hline
\hline
query throughput (qps) & 5,400 & 3,700  \\
\hline
update rate (records per second) & 2,500 & 3,000  \\
\hline
\end{tabular}
\end{center}
\vspace{-0.1in}
\caption{Comparison of Navitas and PDNS background update rate under normal query load}
\vspace{-0.2in}
\label{tab:update}
\end{table}
\vspace{-0.05in}

The background record update rates at a normal server operation point for Navitas and PDNS are listed in Table~\ref{tab:update}. The results are likely sufficient in general ENUM deployment. Furthermore, update performance and query performance balance can be offset by running slaves to take queries and the master to take updates.

\vspace{-0.09in}
\section{Related Work} \label{sec:related}
\vspace{-0.02in}

Benchmarking of ENUM server has been reported recently by Nominum Inc.~\cite{enumbm}. The Nominum work tested ANS, BIND, PDNS-BIND and DJBDNS. Their testbed is also based on Enterprise Linux and they used queryperf load generator as well. However there are significant differences in their tests from ours. For example, Nominum tested PDNS with its BIND backend, citing response time considerations. All their tests for PDNS failed because PDNS-BIND scales badly. We found that response time is not a real concern on our platform and chose to evaluate PDNS with its more scalable MySQL backend. We also instrumented the PDNS server code to examine its internal components, while \cite{enumbm} treats all servers as black box. Overall, we used a much more comprehensive testing methodology and investigated a richer set of performance metrics.

\vspace{-0.05in}
\section{Conclusion} \label{sec:conclusion}

\vspace{-0.02in}
We evaluated ENUM server performance provided by three name server implementations, PDNS, BIND and Navitas. Results show that the query response time on our platform has always been on the order of a few milliseconds or less. Given the several-second-long budget of normal PSTN post dial delay, this latency is not a concern. The query throughput is therefore our key consideration. Since the throughput of BIND for existing records degrades linearly as the record set size grows, BIND is not qualified as a serious ENUM server. Both Navitas and PDNS can serve ENUM, but Navitas outperforms PDNS in all aspects we have tested. Under our 5M-record set test, Navitas achieves six times higher maximum throughput for existing records and an order of two magnitudes higher maximum throughput for non-existing records than basic PDNS does. Navitas also maintains its throughput lead at the presence of background database update load. Our PDNS server profiling reveals that database lookup time is the main cause that constrains PDNS performance, particularly for non-existing records. We studied how PDNS caching can alleviate the problem. By enabling full packet caching, as well as modifying its cache maintenance mechanism, the PDNS throughput for existing records can double. The PDNS throughput for non-existing records is also remarkably improved, but the result is still an order of magnitude lower than the corresponding value of Navitas. The high performance of Navitas may be attributed to its well engineered architecture and optimized implementation. PDNS has at least three areas that can be improved in order to better serve ENUM: implementing a more efficient cache maintenance algorithm; optimizing the memory usage for ENUM records; enhancing the handling of non-existing records by possibly adopting a processing fast path.

\vspace{-0.06in}

\section*{Acknowledgment}

We thank FiberNet Telecom Group for providing the server hardware, Navitas software and other logistics support, as well as assisting in testbed setup. Special thanks to Douglas Kelly, Steve Cespedes, Frank Arundell, Ernest Hoffmann, and Jon Deluca. We also thank Shawn Puliotte and Brian Wellington of Nominum for helpful comments and clarifications.
\bibliographystyle{IEEEtran}
\bibliography{enumperf}
\end{document}